\documentclass[a4paper,
               biblatex,     
               ]{jacow}

\usepackage{pdfpages,multirow,ragged2e} %
\makeatletter%
	\ifboolexpr{bool{xetex}}
	 {\renewcommand{\Gin@extensions}{.pdf,%
	                    .png,.jpg,.bmp,.pict,.tif,.psd,.mac,.sga,.tga,.gif,%
	                    .eps,.ps,%
	                    }}{}
\makeatother

\ifboolexpr{bool{xetex} or bool{luatex}} 
 {}                                      
 {\usepackage[utf8]{inputenc}}           

\usepackage[USenglish]{babel}

\DeclareSIUnit\bar{bar} 

\ifboolexpr{bool{jacowbiblatex}}%
 {
  \addbibresource{MOPA121.bib}
 }{}
\begin{document}

\title{Plasma Lens Prototype Progress: Plasma Diagnostics and Particle Tracking for ILC \NoCaseChange{e\textsuperscript{+}} Source \\[1ex] \small Talk presented at the International Workshop on Future Linear Colliders (LCWS2023), 15-19 May 2023. C23-05-15.3.}

\author{M. Formela\thanks{manuel.formela@desy.de}\textsuperscript{1, 2}, N. Hamann\thanks{niclas.hamann@desy.de}\textsuperscript{1, 2}, G. Moortgat-Pick\textsuperscript{1, 2}, G. Loisch\textsuperscript{1}, K. Ludwig\textsuperscript{1}, J. Osterhoff\textsuperscript{1} \\ \textsuperscript{1} II. Institute of Theoretical Physics, University of Hamburg \\ Luruper Chaussee 149, 22761 Hamburg, Germany \\
\textsuperscript{2} Deutsches Elektronen-Synchrotron DESY, Notkestr. 85, 22607 Hamburg, Germany} 
	
\maketitle

\begin{abstract}
In recent years, the concept of high-gradient, symmetric focusing using active plasma lenses has regained notable attention owing to its potential benefits in terms of compactness and beam dynamics when juxtaposed with traditional focusing elements. An enticing application lies in the optical matching of extensively divergent positrons originating from the undulator-based ILC positron source, thereby enhancing the positron yield in subsequent accelerating structures.
Through a collaboration between the University of Hamburg and DESY Hamburg, a scaled-down prototype for this purpose has been conceptualized and fabricated. In this presentation, we provide an overview of the ongoing progress in the development of this prototype. Furthermore, first insights into the development of a particle tracking code especially designed for plasma lenses with implemented Bayes optimization, are given.
\end{abstract}

\section{INTRODUCTION}
The International Linear Collider (ILC) is a planned electron-positron linear collider designed to operate with energies up to 500\,GeV. To generate positrons, undulator radiation will be directed onto a rotating wheel made of Titanium. This collision setup will facilitate precise measurements of Higgs boson properties, aiming to enhance luminosity and consequently increase the success of the intended collision experiments.
Due to the substantial divergence of produced positrons, the implementation of an optical matching device (OMD) becomes essential to capture the maximum number of positrons while significantly reducing their divergence for downstream accelerator components. This underscores the need to position the OMD as close as feasible to the target. Presently, the Quarter Wave Transformer stands as the favored solution for the OMD. Despite the potential to yield more positrons, the flux concentrator is considered to be unsuitable for use at the ILC positron source, due to the varying focusing field over the ILC's 1\,ms long bunch trains. 
More recently, a novel alternative involving the utilization of current-carrying plasma has emerged, the active plasma lens (APL). In this approach, a plasma is generated by ionizing a gas column and a high-amplitude current pulse is directed through the plasma, inducing azimuthal magnetic fields that exert a radial force on a beam traversing the column.
While the APL holds the promise of significant advantages, it also presents several challenges that must be addressed \cite{Formela:2022gco}. The University of Hamburg, in collaboration with DESY Hamburg, has initiated a project to explore the possibilities and limitations of using active plasma lenses for this purpose. Initial particle tracking simulations have already been conducted to identify the optimal plasma lens design in terms of capturing positrons.
Further research and development of this design are now necessary, encompassing both experimental work with a prototype setup and corresponding simulations that model the hydrodynamics of the current-carrying plasma and the resultant magnetic field. The ongoing status of the prototype design is detailed in the subsequent sections.

\section{PARTICLE TRACKING SIMULATIONS}
In order to investigate the attainable positron yield and determine the operational parameters of the plasma lens, particle tracking simulations have been executed utilizing ASTRA \cite{Floettmann}.

\subsection{Design Optimization}
\begin{figure}[h!]
    \centering
    \includegraphics[scale=0.25]{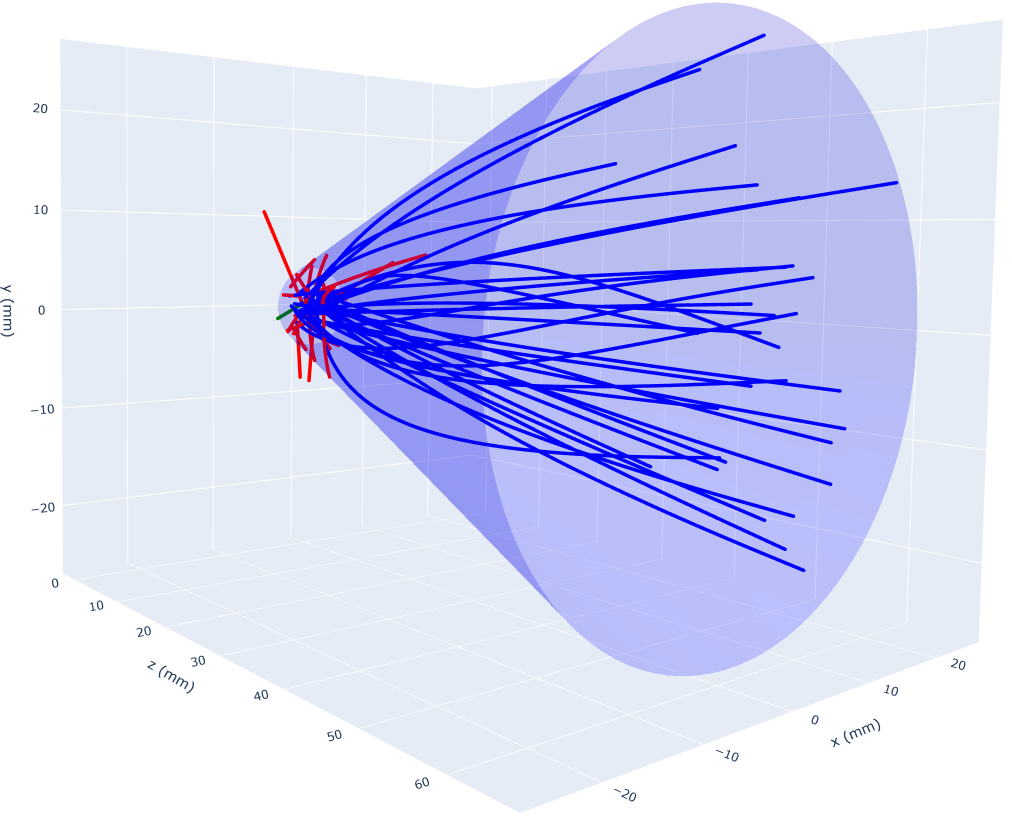}
    \caption{Particle tracking of 54 positrons through the full-scale plasma lens up to a z-position of 65\,mm.}
    \label{fig:track54}
\end{figure}

In order to enhance the plasma lens design for optimal positron yield, the following parameters were subjected to variation: total electric current $I_0$, opening radius $R_0$, exit radius $R_1$, APL length $l$ and the tapering order $n$, which defines the shape of the plasma lens.
This optimization endeavor resulted in a specific plasma lens design that serves as a reference for prototype development.
This optimized design boasts a capture efficiency of approximately $\boldsymbol{43\,\%}$ encompassing a total of 42917 simulated positrons. This efficiency marks an advancement of around twofold in comparison to the current ILC solution \cite{Fukuda2019}, with the positron distribution being provided by M. Fukuda of KEK \cite{Fukuda}. While higher currents are inclined to yield greater efficiencies, this design represents a balanced compromise between capture efficiency and technical feasibility.
An illustrative example employing 54 simulated positrons is presented in Figure \ref{fig:track54}, shedding light on the impact of the plasma lens.

\subsection{Design Stability}
The complete-scale plasma lens design underwent successful stability testing of its capture efficiency by introducing single-parameter errors of $\pm10\,\%$ into the design.
These parameter deviations resulted in a maximum relative reduction of captured particles by up to $\qty{5.0}{\percent}$ \cite{Formela:2022gco}.
Consequently, stability concerns are not anticipated to arise.
However, it's crucial to recognize that any deviation in the peak electric current primarily impacts the capture efficiency.

\section{GAS FLOW SIMULATIONS}
In \cite{hamann:mopa121}, the process of gas flow simulations was described. The insights gained revealed that, on one hand, a balance needs to be struck between the diameter (which determines the strength of gas distribution) and the angle (which determines the gas pressure intensity) of the gas inlets. Additionally, it was found that the gas inlets at the exit of the plasma lens are also suitable for modulating the gas pressure at the inlet. As a result, the gas inlets at the inlet were omitted, and a design featuring two 3\,mm inlets at the exit of the plasma lens with an angle of 70 degrees was agreed upon.
The parameters of the prototype plasma lens can be seen in Table~\ref{tab:optimal_downscaled_design}.

\begin{table}[h!]
\centering
    \caption{Parameters of the Prototype Plasma Lens Design}
   \begin{tabular}{lccc}
       \toprule
       \textbf{Parameter name} & \textbf{Symbol} & \textbf{Unit} & \textbf{Value} \\
       \midrule
    Total Electric Current & $I_0$ & $\unit{\A}$ & $\num{350}$ \\
    Tapering Type & & & linear \\
    Opening Radius & $R_\mathrm{0}$ & $\unit{\mm}$ & $\num{0.85}$ \\
    Exit Radius & $R_\mathrm{1}$ & $\unit{\mm}$ & $\num{5}$ \\
    Tapering Length & $L$ & $\unit{\mm}$ & $\num{12}$ \\
       \bottomrule
   \end{tabular}
   \label{tab:optimal_downscaled_design}
\end{table}

\section{PROTOTYPE DEVELOPMENT}
\begin{figure}[h!]
    \centering
    \includegraphics[scale=0.2]{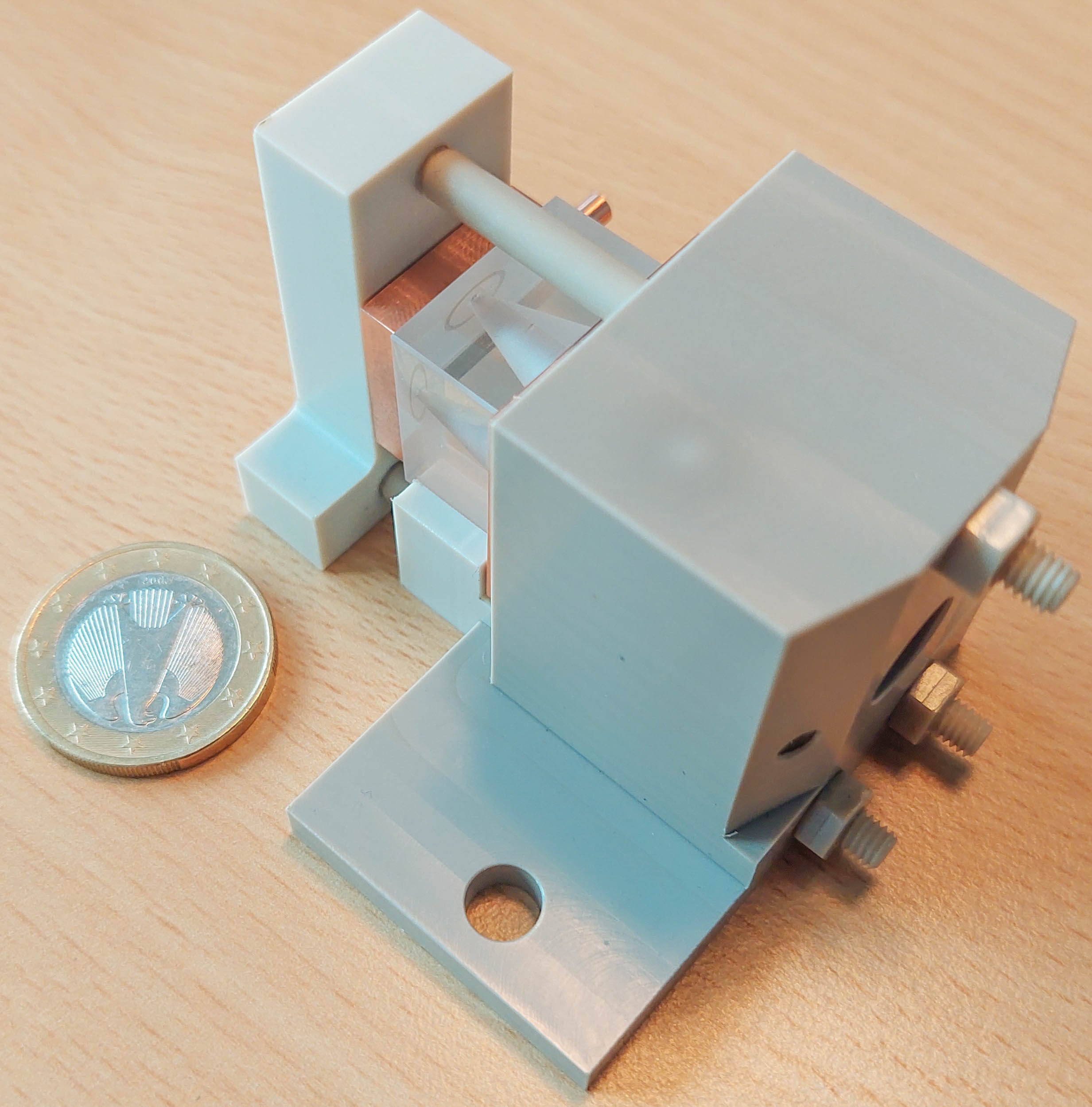}
    \caption{Image of the prototype plasma lens set-up.}
    \label{fig:image}
\end{figure}

The ADVANCE Lab at DESY, where the prototype is slated for testing \cite{Garland:2022lgn}, imposes a peak current limitation of 350\,A.
To achieve the same current density using 350\,A instead of 9\,kA, the prototype has been scaled down by a factor of 5.07 \cite{Formela:2022gco}.
Based on the outcomes of the gas flow simulations, a design has been formulated using Autodesk Inventor.
This configuration comprises three mounting brackets crafted from Polyetheretherketone (PEEK), and two electrodes fashioned from low oxygen copper.
Assembly is achieved through customised threaded PEEK rods and nuts, while sealing is accomplished with o-rings positioned between the plasma lens, electrodes and mounting brackets.
The mounting brackets serve the purpose of securing the plasma lens and electrodes in place while also integrating the gas inlets.
To accommodate the angled gas inlets, the mounting bracket edges at the exit of the plasma lens have been bevelled.
Additionally, all dimensions have been adapted to align with the mounting table within the vacuum tank, ensuring precise measurements.
Figure \ref{fig:image} illustrates an image of the completed prototype setup.
All components were fabricated in the University workshop at DESY.

\section{FIRST TEST RESULTS}
\begin{figure}[h!]
    \centering
    \includegraphics[scale=0.15]{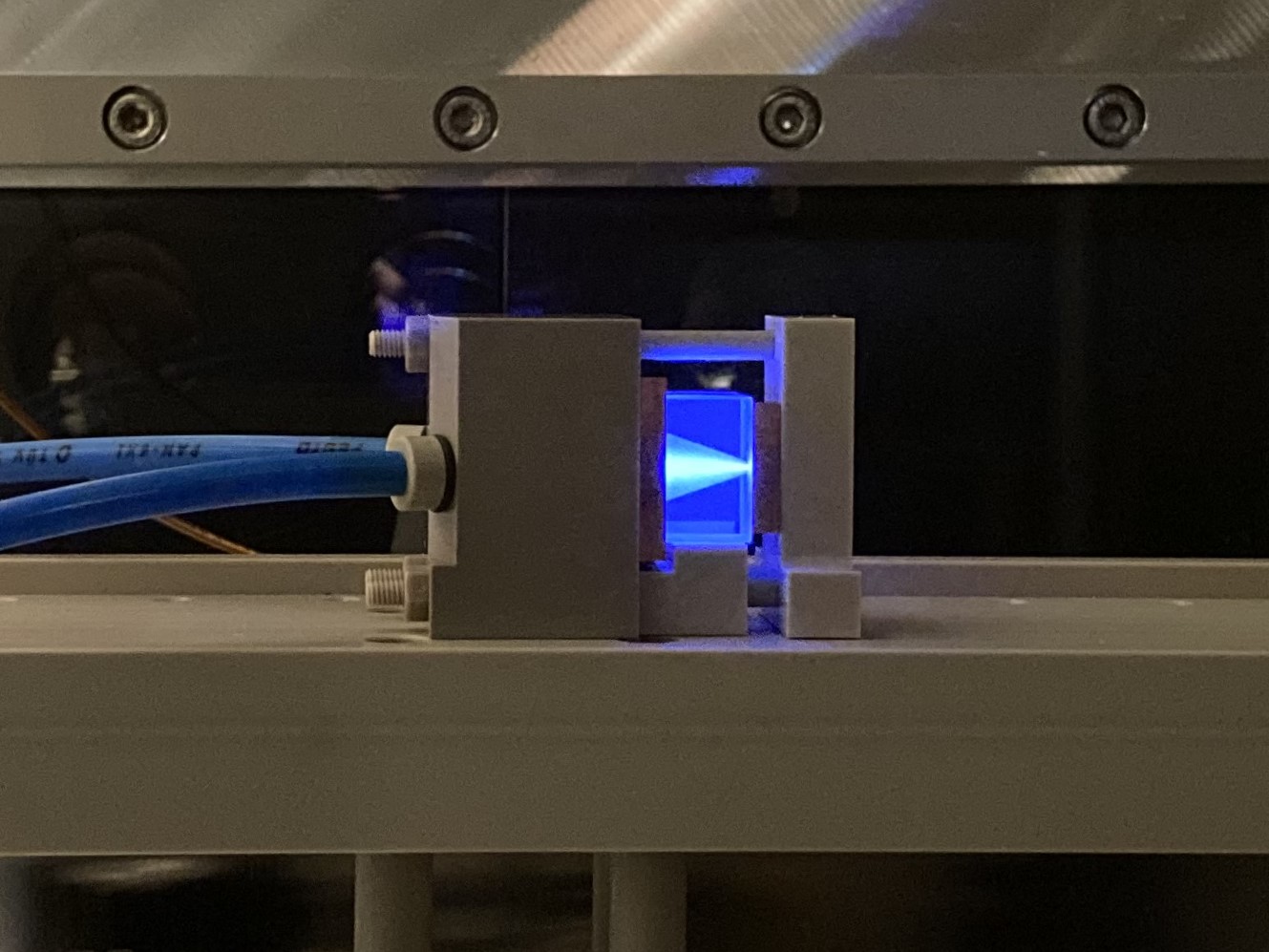}
    \caption{Produced plasma in the prototype in the ADVANCE LAB at DESY.}
    \label{fig:prod_plasma}
\end{figure}

After the prototype was completed, it was subjected to an initial test in the vacuum chamber at the ADVANCE LAB at DESY. Initially, no gas discharge could be ignited within the plasma lens. However, as the voltage was increased, plasma was successfully generated within the lens for the first time. A few images and current curves were recorded. In Figure~\ref{fig:prod_plasma}, it can be observed that the current distribution within the lens is not as idealized as initially assumed. Instead of filling the entire cone, the plasma only occupies a small portion. Furthermore, observations from various discharges indicate that the plasma is not stable and thus not reproducible. This instability is, of course, unfavorable for application in the ILC positron source.
The initial hypothesis regarding the cause is that the plasma is pinching. This occurs when the pressure of the plasma is lower than the magnetic pressure, which compresses the current through the induced magnetic field. This pressure drop is generated here by increasing the radius of the lens. However, further measurements are necessary to better understand this behavior.

\section{TRACKING CODE DEVELOPMENT}
In addition to prototype development, work is being done on a tracking code for plasma lenses. This code is based on Newtonian mechanics and the Lorentz force and is written in Python 3.10. The equations of motion are determined approximately by simply integrating velocity and doubly integrating position. By making the step size smaller, the solution converges. In its current state, the code only consists of the plasma lens, but additional elements from the positron injection scheme are currently being implemented, namely the 0.5\,T strong solenoid and the first standing wave tube. Since the previous tracking simulations were done using ASTRA, this code is based on it. This means that the underlying particle distribution for this code is the same as for ASTRA. The idea of controlling the simulation through an initialization file with a .in extension has also been adopted from ASTRA. Therefore, ultimately, as long as Python and all relevant packages have been correctly installed, no additional terminal needs to be opened. Once the code is complete with all its components, there will be comprehensive documentation. Currently, it is only possible to work with electrons and positrons. Additional particle species, such as protons, are planned to be included later. The plasma lens can be modified by several parameters, with the most significant being peak current, length, aperture radius, and exit radius. Currently, it's possible to choose between a uniform and a non-uniform current distribution according to \cite{Tilborg} and \cite{Roeckemann}. It's also feasible to utilize a self-simulated magnetic field. For now, only the counts of active and passive particles are recorded after the simulation. Once the solenoid and SWT have been integrated, there will also be an option for the longitudinal cut, which was used in ASTRA simulations to simulate the energy acceptance of the damping ring. In addition to the tracking code, there is already a 3D plot code based on Plotly. This allows tracking the paths of individual particles in 3D. Figure~\ref{fig:track54} was created with the tracking code and the 3D plot code. Another aspect is the implementation of a Bayesian optimization algorithm via Botorch. This is intended to make it possible in the future to optimize the number of active particles after the longitudinal cut with a given particle distribution, and thus provide an idea of how the plasma lens could be shaped using parameter sets.

\section{SUMMARY AND OUTLOOK}
In a collaborative effort between the University of Hamburg and DESY Hamburg, a project has been initiated to assess the viability of utilizing an active plasma lens as an optical matching component for the ILC's undulator-based positron source. Particle tracking simulations have been executed to establish an initial parameter configuration for the plasma lens, ensuring a substantial positron yield.
To alleviate the demands on high voltage drive electronics, a prototype on a reduced scale has been constructed. Gas flow simulations have been conducted to ascertain the appropriate geometry for the prototype setup.
First tests were conducted at the ADVANCE LAB at DESY \cite{Garland:2022lgn}. After a few attempts, it was possible to generate plasma. However, this plasma is likely pinching and is not stable. The next step involves capturing the discharge with high temporal resolution to identify the parameters at which the plasma starts pinching. The particle tracking code is being further developed and is intended to enable proton tracking in the future, as well as accommodate the solenoid and the standing wave tube situated behind the plasma lens.
In parallel, further parameter optimisation and magnetohydrodynamics simulations of the plasma will be performed. 

\section{ACKNOWLEDGEMENTS}

The authors would like to thank S. Riemann, M. Fukuda, T. Parikh, and J. Garland for useful discussions. %

%
%

\ifboolexpr{bool{jacowbiblatex}}%
	{\printbibliography}%
	

\end{document}